\documentclass{article}
\usepackage{amsmath}
\title{A Fast Parametric Ellipse Algorithm}
\author{Jerry R. Van\nobreak\hspace{.11em}Aken}
\usepackage{graphicx}
\usepackage{hyperref}
\usepackage{lmodern}
\usepackage[T1]{fontenc}
\usepackage{relsize}
\usepackage{amssymb}
\usepackage{cprotect}
\usepackage{mathtools}
\usepackage[title]{appendix}
\mathchardef\mhyphen="2D

\begin{document}
  \maketitle

\begin{abstract}
This paper describes a 2-D graphics algorithm that uses shifts and adds to precisely plot a series of points on an ellipse of any shape and orientation. The algorithm can also plot an elliptic arc that starts and ends at arbitrary angles. The ellipse algorithm described here is largely based on earlier papers by Van\nobreak\hspace{.11em}Aken and Simar [1,\,2]. A new flatness test is presented for automatically controlling the spacing between points plotted on an ellipse or elliptic arc. Most of the calculations performed by the ellipse algorithm and flatness test use fixed-point addition and shift operations, and thus are well-suited to run on less-powerful processors. C++ source code listings are included. \\

\noindent \emph{Keywords:} parametric ellipse algorithm, rotated ellipse, Minsky circle algorithm, elliptic arc, flatness, conjugate diameters, affine invariance 
\end{abstract}

\section{Introduction}

This paper describes a 2-D graphics algorithm that uses fixed-point addition and shift operations to precisely plot a series of points on an ellipse of any shape and orientation. The ellipse algorithm is largely based on earlier papers by Van\nobreak\hspace{.11em}Aken and Simar [1,\,2], which extend Marvin Minsky's well-known circle algorithm [3,\,4,\,5] to ellipses, and show how to cancel out the sources of error in Minsky's original algorithm. The ShapeGen graphics library\footnote{ShapeGen is an open-source 2-D graphics library. The C++ source code and documentation are available at https://www.github.com/jvanaken1/shapegen.} [6] uses the resulting ellipse algorithm to generate circles, ellipses, elliptic arcs, and elliptic splines.

The ellipse algorithm significantly reduces the processing time needed to calculate each point on an ellipse or elliptic arc. The algorithm's inner loop requires only four additions and four right-shift operations to calculate a point on an ellipse centered at the $x$-$y$ coordinate origin. Two more addition operations are required to translate the ellipse center to a position other than the origin. Fixed-point arithmetic is used. The points plotted by the ellipse algorithm fall precisely on the ellipse, to within the precision afforded by the fixed-point representation.

A friendly interface enables the graphics user to specify an ellipse in terms of the square, rectangle, or parallelogram in which it is inscribed. The ellipse's center coincides with the center of the enclosing parallelogram. The ellipse touches (and is tangent to) the parallelogram at the midpoint of each of its four sides.

In fact, the ellipse algorithm described here requires only three points to completely specify the ellipse: the center point, and the end points of a pair of \emph{conjugate diameters} of the ellipse [1,\,2,\,7]. As shown in \mbox{Figure 1}, the two conjugate diameter end points, labeled $P$ and $Q$, are simply the midpoints of two adjacent sides of the enclosing parallelogram. Conversion between the parallelogram and the three points describing the ellipse is simple enough that these two representations are essentially equivalent.

\begin{figure}
\advance\leftskip-0.2cm
\includegraphics[width=355pt]{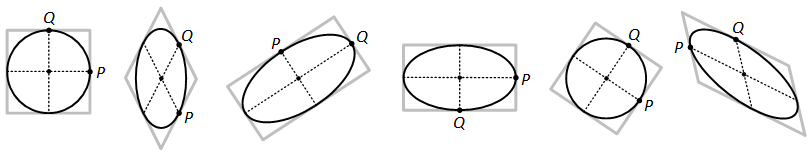}
\caption{The end points $P$ and $Q$ of a pair of conjugate diameters of an ellipse are simply the midpoints of two adjacent sides of the square, rectangle, or parallelogram in which the ellipse is inscribed.}
\end{figure}

In some applications, the direction of rotation in which the ellipse or elliptic arc is drawn is important. To draw a full ellipse, the ellipse algorithm plots points starting at the first conjugate diameter end point, $P$, and moving toward the second, $Q$. When drawing an elliptic arc, the arc's starting angle is measured relative to $P$, and is positive in the direction of $Q$. The arc's sweep angle is positive in the same direction. Either angle can be positive or negative. 

The ShapeGen graphics library [6] renders an elliptic arc or other curve by plotting a series of points at more\:\!-or-less regular intervals along the curve, and then connecting each pair of adjacent points on the curve with a straight line segment. The library user controls the spacing between successive points on an ellipse or other curve by specifying a \emph{flatness} parameter (Adobe [8] p. 181). The process of approximating a curve with a series of chords or polygonal edges is called \emph{flattening} ([8] p. 502). The flatness parameter is the maximum error tolerance of a flattened representation, as specified by the user, and is defined to be the largest gap, in pixels, that is permitted between a chord and the curve segment that it represents. Flatness control for ellipses and elliptic arcs is discussed in detail in a later section.

\section{Parametric equations of an ellipse}

In this section, we will derive the parametric equations of an ellipse, given the end points, $P$ and $Q$, of a pair of conjugate diameters of the ellipse. To simplify the calculations, the ellipse is centered at the origin. 

\begin{figure}
\centering
\includegraphics[width=300pt]{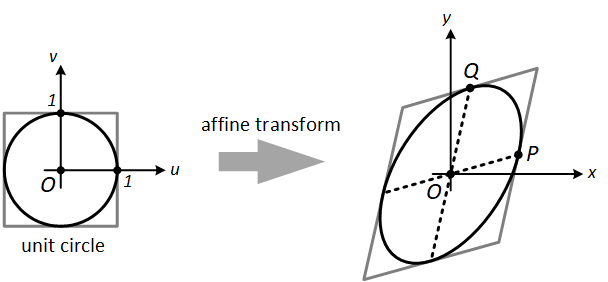}
\caption{Affine transformation of a unit circle and its bounding square to an ellipse and its bounding parallelogram.}
\end{figure}

\mbox{Figure 2} shows that an ellipse and its bounding parallelogram are the affine\:\!-transformed images of a unit circle and its bounding square. We use $u$-$v$ coordinates for the unit circle, and $x$-$y$ coordinates for the ellipse. Points $(1,0)$ and $(0,1)$ on the unit circle are transformed to points $P = (x_P,y_P)$ and $Q = (x_Q,y_Q)$ on the ellipse. Any pair of perpendicular diameters of the circle are conjugate diameters of the circle. An affine transformation of the circle to an ellipse maps a pair of conjugate diameters of the circle to conjugate diameters of the ellipse. Thus, $P$ and $Q$ in \mbox{Figure 2} are end points of conjugate diameters of the ellipse.

The affine transformation of a point $(u,v)$ on the unit circle to a point $(x,y)$ on the ellipse can be expressed as
\begin{align*}
\begin{bmatrix}
x \\  y 
\end{bmatrix}
= \mathbf{M} 
\begin{bmatrix}
u \\  v 
\end{bmatrix}
\quad\:\textrm{ where }\:\mathbf{M} =
\begin{bmatrix}
m_{11}  & m_{12}  \\ m_{21}  & m_{22}  
\end{bmatrix}
\end{align*}
where the circle and ellipse are centered at their respective origins. To find the coefficients $m_{ij}$ of transformation matrix $\mathbf{M}$, consider that conjugate diameter end points $P$ and $Q$ on the ellipse are related to points $(1,0)$ and $(0,1)$ on the unit circle by the expressions $P = \mathbf{M} \mathsmaller{\begin{bmatrix} 1 & 0 \end{bmatrix}^{\mathsf{T}} }$ and $Q = \mathbf{M} \mathsmaller{\begin{bmatrix} 0 & 1 \end{bmatrix}^{\mathsf{T}} }$. These two expressions expand to the following:
\begin{align*}
\\[-14pt]
\begin{bmatrix}
x_P \\ y_P
\end{bmatrix}
& =
\begin{bmatrix}
m_{11} & m_{12} \\
m_{21} & m_{22}
\end{bmatrix}
\begin{bmatrix}
1 \\ 0
\end{bmatrix}   \\[+2pt]
& =
\begin{bmatrix}
m_{11} \\ m_{21}
\end{bmatrix}
\\[+8pt]
\begin{bmatrix}
x_Q \\ y_Q
\end{bmatrix}
& =
\begin{bmatrix}
m_{11} & m_{12} \\
m_{21} & m_{22} 
\end{bmatrix}
\begin{bmatrix}
0 \\ 1
\end{bmatrix}   \\[+2pt]
& =
\begin{bmatrix}
m_{12} \\ m_{22}
\end{bmatrix}
\end{align*}
From inspection, we see that
\begin{equation*}
\mathbf{M} =
\begin{bmatrix}
x_P & x_Q \\
y_P & y_Q 
\end{bmatrix}
\end{equation*}
The affine transformation of a point $(u,v)$ on the unit circle to a point $(x,y)$ on the ellipse can now be expressed as
\begin{align}
\begin{bmatrix}
x \\ y
\end{bmatrix}
& =
\begin{bmatrix}
x_P & x_Q \\
y_P & y_Q 
\end{bmatrix}
\begin{bmatrix}
u \\ v
\end{bmatrix}
\end{align}
where the matrix coefficients are the $x$ and $y$ coordinates of end points $P$ and $Q$ of a pair of conjugate diameters of the ellipse.

The parametric equations for the unit circle are
\begin{align}
u(\theta) & = \cos{\theta} \\
v(\theta) & = \sin{\theta} \nonumber
\end{align}
for $0 \le \theta \le 2\pi$. The corresponding parametric equations for the ellipse (Foley \emph{et al} [9]) can be obtained by substituting equations (2) into \mbox{equation (1)}, which yields\footnote{Given conjugate diameter end points $P = (x_P,y_P)$ and $Q = (x_Q,y_Q)$ on an ellipse, equations (3) can be used to generate additional conjugate diameter end points $P' = (x(\theta),y(\theta))$ and $Q' = (x(\theta\pm\frac{\pi}{2}),y(\theta\pm\frac{\pi}{2}))$ on the same ellipse. Also, see equations (15) and (16).}  \\[-20pt]
\begin{align}
x(\theta) & = x_P \cos{\theta} + x_Q \sin{\theta} \\
y(\theta) & = y_P \cos{\theta} + y_Q \sin{\theta} \nonumber
\end{align}
for $0 \le \theta \le 2\pi$.

These equations can also be expressed in the following sequential form, which can be used to plot a series of $N$ points on the ellipse by increasing angle $\theta$ from $0$ to $2 \pi$ radians in incremental steps of size $\alpha = 2 \pi / N$ radians: 
\begin{align}
x_n & = x_P \cos(n \alpha) + x_Q \sin(n \alpha)    \\
y_n & = y_P \cos(n \alpha) + y_Q \sin(n \alpha)
\end{align}
for $n = 1,2,...,N$, where $P = (x_P,y_P)$ and $Q = (x_Q,y_Q)$ are end points of a pair of conjugate diameters of the ellipse. However, directly using these equations requires evaluating the sine and cosine terms in the inner loop of the ellipse algorithm, which could be slow.

\section{A faster inner loop}

To speed up the inner loop of the ellipse algorithm, we will adapt a well-known circle\:\!-generating algorithm that was discovered by Marvin Minsky in the early 1960s (Beeler \emph{et al} [3], Paeth [4], Ziegler Hunts \emph{et al} [5]). This circle generator can rotate a point about the origin using only shifts and adds. Here's the inner loop of the algorithm:\\[-20pt]
\begin{align}
u_n & = u_{n-1} - \varepsilon \!\: v_{n-1}    \\
v_n & = \varepsilon \!\: u_n + v_{n-1} 
\end{align}
where $0 < \varepsilon \le 1$, and $1 \le n \le \lfloor 2 \pi / \varepsilon \rfloor$. We use $u$-$v$ coordinates here to avoid confusion with the $x$-$y$ coordinates used for the ellipse in equations (4) and (5).

The circle generator initially sets the $u$ and $v$ values to a starting point $(u_0, v_0)$ on a circle centered at the origin. Each iteration of equations (6) and (7) plots the next point on the (approximate) circle by rotating the previous point about the center by an (approximate) angular increment of $\varepsilon$ radians. The circle generator draws (in Minsky's words [3]) an ``almost circle''---a nearly circular ellipse---and the ellipse becomes a better approximation to a circle as the size of angular increment $\varepsilon$ is decreased.

The angular increment can be set to a negative power of two ($\varepsilon = 1/2^k$ for $k = 0,1,2,...$), for which the multiplications in equations (6) and (7) can be replaced with simple integer right-shift operations. With this modification, fixed-point arithmetic can be used to perform the inner-loop calculations. For plotting points on a typical graphics display, the $u$ and $v$ values in these equations can be represented as 16.16 fixed-point values; these values are stored as signed 32-bit integers, but the 16 least-significant bits are assumed to lie to the right of the binary point, and represent fractional values.

The circle generator looks similar to a standard 2-D rotation\footnote{A standard rotation has the form $\begin{bmatrix} u_n \\  v_n \end{bmatrix} =  \begin{bmatrix}\cos \alpha & -\sin \alpha \\ \sin \alpha & \sin \alpha \end{bmatrix} \begin{bmatrix} u_{n -1} \\ v_{n-1} \end{bmatrix}$.} that uses the approximations $\cos \varepsilon \approx 1$ and $\sin \varepsilon \approx \varepsilon$, which are useful approximations if the angular increment $\varepsilon$ is small. However, the calculation of $v_n$ in equation (7) uses $u_n$ rather than $u_{n-1}$. Although this subscript value might look like a typing error, it is actually key to the algorithm's behavior (Newman and Sproull [10], Blinn [11]). The result is that the determinant of equations (6) and (7) is unity. This fact can be verified by substituting the right-hand side of equation (6) in place of $u_n$ in equation (7) and then expressing the equations in matrix form, as follows: \\[-20pt]
\begin{align*}
\begin{bmatrix}
u_n \\ v_n
\end{bmatrix}
& =
\begin{bmatrix}
1 & -\varepsilon \\
\varepsilon & 1-\varepsilon^2
\end{bmatrix}
\begin{bmatrix}
u_{n-1} \\ v_{n-1}
\end{bmatrix}
\end{align*}
Because the determinant is unity, the curve plotted by Minsky's circle generator closes as it completes a full rotation. In fact, the circle generator is remarkably stable---if the inner loop is allowed to free\:\!-run for many rotations, it will continue to plot points on the same approximate circle without spiraling inward or outward.

The major drawback to using Minsky's circle generator is that it draws only approximate circles, whereas we want an algorithm to draw ellipses precisely. To fix this problem, we will identify the sources of error in the circle generator and cancel them out.

Analyses by Van\,Aken and Simar [1,\:2] and Ziegler Hunts \emph{et al} [5] found that the result of $n$ iterations of equations (6) and (7) is as follows:
\begin{equation}
u_n = u_0 \cos(n \alpha) - \left[ \frac{v_0 - \frac{\mathlarger{\varepsilon}}{2} u_0}{\sqrt{1 - \frac{1}{4} \varepsilon^2 }} \right] \sin(n \alpha)   \\
\end{equation}
\begin{equation}
v_n = \left[ \frac{u_0 - \frac{\mathlarger{\varepsilon}}{2} v_0}{\sqrt{1 - \frac{1}{4} \varepsilon^2 }} \right] \sin(n \alpha) + v_0 \cos(n \alpha)
\end{equation}
where $(u_0,v_0)$ is the starting point on the circle, and $\alpha$ is the precise angular increment, to which the value $\varepsilon$ in equations (6) and (7) is only an approximation.
Angular increment $\alpha$ in equations (8) and (9) is related to $\varepsilon$ by the expressions
\begin{align}
\sin \frac{\alpha}{2} = \frac{\varepsilon}{2} \qquad \textrm{and} \qquad
\cos \frac{\alpha}{2} = \sqrt{1 - \mathsmaller{\frac{1}{4}} \varepsilon^2 }
\end{align}
For a detailed analysis of the circle generator, see \mbox{Appendix A}.

To precisely plot points on a circle, the messy-looking terms in brackets in equations (8) and (9) should equal $v_0$ and $u_0$, respectively. The terms in brackets can be made to approach the desired values, $v_0$ and $u_0$, by decreasing the size of $\varepsilon$, but we would prefer a solution that wholly eliminates the error for all values of $\varepsilon$.

The error in either equation (8) or (9) can be eliminated by substituting an altered initial value in place of either $v_0$ or $u_0$. For example, the error in equation (9) can be eliminated by replacing $u_0$ with a value $U_0$ specified so that
\begin{align*}
u_0 = \frac{U_0 - \frac{\mathlarger{\varepsilon}}{2} v_0}{\sqrt{1 - \frac{1}{4} \varepsilon^2 }}
\end{align*}
Solving for $U_0$, we have
\begin{equation}
U_0 = u_0 \sqrt{1 - \mathsmaller{\frac{1}{4}} \varepsilon^2 } \: + \: \frac{\varepsilon}{2}  v_0
\end{equation}
Substituting $U_0$ for $u_0$ in equations (8) and (9) yields
\begin{align}
u_n &  = \Big( u_0 \sqrt{1 - \mathsmaller{\frac{1}{4}} \varepsilon^2 } \: + \: \frac{\varepsilon}{2}  v_0 \Big) \cos(n \alpha) - \Big( v_0 \sqrt{1 - \mathsmaller{\frac{1}{4}} \varepsilon^2 } \: - \: \frac{\varepsilon}{2} u_0 \Big) \sin(n \alpha)  \\[+4pt]
v_n & = v_0 \cos(n \alpha) + u_0 \sin(n \alpha)
\end{align}
Equation (13) shows that the circle generator can precisely perform the calculations required in equations (4) and (5) by using fixed-point shifts and adds in place of floating-point sine and cosine functions.

\section{Ellipse algorithm}

By comparing equation (13) with equations (4) and (5), we see that the ellipse algorithm will need to incorporate two copies of the circle generator. The first copy will produce $x_n$ in equation (4), and the second will produce $y_n$ in \mbox{equation (5)}.

The following C++ function\cprotect\footnote{This implementation of the \verb|CircleGen| function assumes that the C++ compiler supports arithmetic right-shift, as is required by the C++20 standard.} implements the circle generator in equations (6) and (7), and is called twice in the ellipse algorithm's inner loop:
\begin{verbatim}
    inline void CircleGen(FIXED& u, FIXED& v, int k)
    {
        u -= v >> k;
        v += u >> k;
    }
\end{verbatim}
The parameters of type \texttt{FIXED} in this listing are 16.16 fixed-point values; as previously discussed, these values are stored as signed 32-bit integers, but the 16 least-significant bits are assumed to lie to the right of the binary point, and represent fractional values. Parameters \texttt{u} and \texttt{v} represent the values $u$ and $v$ in equations (6) and (7). The angular increment $\varepsilon$ in these equations is set to a negative power of two, $\varepsilon = 1/2^k$ for $k=0,1,2,...$, where exponent $k$ is specified by parameter \texttt{k}. The \texttt{inline} qualifier means that the \texttt{CircleGen} function incurs no function-call performance penalty.

For equation (11), $U_0$ can be calculated by using floating-point arithmetic and calling the \texttt{sqrt} function in standard C header file \texttt{math.h}. The square-root operation can be avoided by storing the values $\sqrt{1-\frac{1}{4}\varepsilon^2}\,$, where $\varepsilon=1/2^k$, as table entries for $k=0,1,2,...$, but floating-point multiplication is still required. A possibly faster option for less-powerful processors is to use a Taylor series approximation for the square root, in which case $U_0$ can be expressed as
\begin{equation}
U_0 = u_0 \Big( 1 - \frac{1}{8}\varepsilon^2 - \frac{1}{128}\varepsilon^4 - \frac{1}{1024}\varepsilon^6 - \frac{5}{32768}\varepsilon^8 - ... \Big) + \frac{\varepsilon}{2}v_0
\end{equation}
where $\varepsilon = 1/2^k$. In this form, $U_0$ can be calculated using fixed-point arithmetic. The middle three terms inside the parentheses are negative powers of two, and multiplications of these terms by $u_0$ can be performed as right-shift operations. For typical graphics applications, sufficient accuracy is obtained by truncating the Taylor series after the 6th-order term.

The following C++ function uses equation (14) to calculate $U_0$, but omits the 8th-order and higher terms:
\begin{verbatim}
    FIXED InitialValue(FIXED u0, FIXED v0, int k)
    {
        int shift = 2*k + 3;
        FIXED w = u0 >> shift;
        FIXED U0 = u0 - w + (v0 >> (k + 1));
    
        w >>= shift + 1;
        U0 -= w;
        w >>= shift;
        U0 -= w;
        return U0;
    }
\end{verbatim}
Parameters \texttt{u0} and \texttt{v0} represent the initial values, $u_0$ and $v_0$, in equation (14). Approximate angular increment $\varepsilon = 1/2^k$, where exponent $k$ is represented by parameter \texttt{k}. 

The ellipse algorithm calls the \texttt{InitialValue} function twice---the first call adjusts the initial value of the $x_Q$ coordinate, and the second adjusts the initial value of $y_Q$. These adjustments ensure the accuracy of the $x$-$y$ coordinates that are calculated in the algorithm's inner loop and used to plot points on the ellipse.

The following C++ function implements the core ellipse algorithm, and can plot both ellipses and elliptic arcs:
\begin{verbatim}
    void EllipseCore(FIXED xC, FIXED yC, FIXED xP, FIXED yP, 
                     FIXED xQ, FIXED yQ, FIXED sweep, int k)
    {    
        int count = sweep >> (16 - k);

        PlotPoint(xP+xC, yP+yC);
        xQ = InitialValue(xQ, xP, k);
        yQ = InitialValue(yQ, yP, k);
        for (int i = 0; i < count; ++i)
        {
            CircleGen(xQ, xP, k);
            CircleGen(yQ, yP, k);
            PlotPoint(xP+xC, yP+yC);
        }
    }
\end{verbatim}
This function calls the two functions, \texttt{CircleGen} and \texttt{InitialValue}, that were previously discussed. It also calls an inline function, \texttt{PlotPoint}, that we can assume plots a point that's specified by its $x$-$y$ coordinates (in 16.16 fixed-point format). Parameters \texttt{xC} and \texttt{yC} are the $x$-$y$ coordinates at the center of the ellipse. Parameters \texttt{xP} and \texttt{yP} specify the first conjugate diameter end point, $P = (x_P, y_P)$, and \texttt{xQ} and \texttt{yQ} specify the second conjugate diameter end point, $Q = (x_Q, y_Q)$. End points $P$ and $Q$ are specified with \emph{center-relative} coordinates; that is, as $x$-$y$ offsets from the ellipse center. Parameter \texttt{sweep} specifies the sweep angle, which is the angle traversed by the elliptic arc. The \texttt{sweep} parameter is a positive, fixed-point value, and is expressed in radians. The approximate angular increment between points plotted by the \texttt{EllipseCore} function is $\varepsilon = 1/2^k$ radians, where $k$ is specified by parameter \texttt{k}.

An arc plotted by the \texttt{EllipseCore} function always starts at point $P$ and steps in the direction of point $Q$. However, a calling function can set an arbitrary arc starting point on the ellipse, and can support both positive and negative sweep angles. To do so, the caller modifies the conjugate diameter end points before passing them to \texttt{EllipseCore}, as will be described shortly.

To determine how many points to plot, the \texttt{EllipseCore} function divides the sweep angle by angular increment $\varepsilon$ and then truncates the fixed-point result to an integer value. This calculation could be performed by shifting the \texttt{sweep} parameter value left by \texttt{k} bits, and then shifting the result right by 16 bits, but because \texttt{k} will always be less than 16, it's safe to combine these two operations into a single right-shift by $16-\texttt{k}$ bits. A later section will discuss limits on the size of the \texttt{k} parameter in more detail.

The graphics library user doesn't directly call the \texttt{EllipseCore} function to plot ellipses and elliptic arcs. Instead, the user calls two public interface functions, \texttt{PlotEllipse} and \texttt{PlotEllipticArc}, which will be described next, and these functions call \texttt{EllipseCore}. In calls to these two public functions, the user specifies window-relative or viewport-relative coordinates for the ellipse center and the two conjugate diameter end points, $P$ and $Q$. However, the \texttt{PlotEllipse} and \texttt{PlotEllipticArc} functions convert $P$ and $Q$ to center-relative coordinates before passing them to the \texttt{EllipseCore} function\cprotect\footnote{Users frequently transform lists of points to window- or viewport-relative coordinates before passing them to user-callable functions like \verb|PlotEllipse| and \verb|PlotEllipticArc|. For convenience, these functions translate conjugate diameter end points to center-relative coordinates rather than requiring the user to do so.}.

The \texttt{PlotEllipse} function is relatively simple. Here's the C$++$ implementation:
\begin{verbatim}
   void PlotEllipse(FIXED xC, FIXED yC, FIXED xP, FIXED yP,
                    FIXED xQ, FIXED yQ, int k)
   {
      EllipseCore(xC, yC, xP-xC, yP-yC, xQ-xC, yQ-yC, FIX_2PI, k);
   }
\end{verbatim}
The sweep angle passed to the \texttt{EllipseCore} function is set to the constant \verb|FIX_2PI|, which is the quantity $2\pi$ in 16.16 fixed-point format.  

The \texttt{PlotEllipticArc} function is more complex. It plots an arc of an ellipse given the arc's starting angle and its sweep angle. If $P$ and $Q$ are the caller-specified end points of a pair of conjugate diameters on the ellipse, and the caller-specified arc starting angle $\varphi$ is nonzero, the function generates a new pair of conjugate diameter end points, $P'$ and $Q'$, where $P'$ is the arc starting point (as determined by $\varphi$), and $P'$ and $Q'$ define the same ellipse as the original end points, $P$ and $Q$. The \texttt{PlotEllipticArc} function then passes $P'$ and $Q'$ to the \texttt{EllipseCore} function, which uses $P'$ as the arc starting point.

To obtain expressions for points $P'$ and $Q'$, consider that affine transformations preserve the conjugate diameters property. For example, points $P$ and $Q$ on the ellipse in \mbox{Figure 2} must lie on conjugate diameters of the ellipse because they are the transformed images of points $(1,0)$ and $(0,1)$ on the unit circle, which lie on conjugate diameters of the circle. To aid in our discussion, we'll label these points $A = (1,0)$ and $B = (0,1)$. If $A$ and $B$ on the unit circle are simultaneously rotated by starting angle $\varphi$, the rotated points, $A'$ and $B'$, lie on diameters that are perpendicular, and thus are conjugate diameters of the circle. If these two rotated points on the circle are then transformed to points $P'$ and $Q'$ on the ellipse, $P'$ and $Q'$ must lie on conjugate diameters of the ellipse.

First, to obtain the $u$-$v$ coordinates of points $A'$ and $B'$ on the unit circle, points $A$ and $B$ are rotated by arc starting angle $\varphi$, as follows:
\begin{align*}
A' & =
\begin{bmatrix}
\cos{\varphi} & -\sin{\varphi} \\
\sin{\varphi} & \cos{\varphi}
\end{bmatrix}
A   \\
& =
\begin{bmatrix}
\cos{\varphi} \\ \sin{\varphi}
\end{bmatrix}
\\[+6pt]
B' & =
\begin{bmatrix}
\cos{\varphi} & -\sin{\varphi} \\
\sin{\varphi} & \cos{\varphi}
\end{bmatrix}
B   \\
& =
\begin{bmatrix}
-\sin{\varphi} \\ \cos{\varphi}
\end{bmatrix}
\end{align*}
Next, equation (1) is used to transform points $A'$ and $B'$ on the unit circle to points $P' = (x_P',y_P')$ and $Q' = (x_Q',y_Q')$ on the ellipse:
\begin{align*}
P' & =
\begin{bmatrix}
x_P & x_Q \\
y_P & y_Q
\end{bmatrix}
A'   \\[+6pt]
Q' & =
\begin{bmatrix}
x_P & x_Q \\
y_P & y_Q
\end{bmatrix}
B'
\end{align*}
where the matrix coefficients are the $x$ and $y$ coordinates of the original conjugate diameter end points, $P=(x_P,y_P)$ and $Q=(x_Q,y_Q$). Finally, these expressions are expanded, as follows, to obtain the \mbox{$x$-$y$} coordinates of $P' $ and $Q'$ in terms of $P$, $Q$, and arc starting angle $\varphi$:
\begin{align}
x_P' & = x_P \cos{\varphi} + x_Q \sin{\varphi} \\
y_P' & = y_P \cos{\varphi} + y_Q \sin{\varphi} \nonumber \\[+8pt]
x_Q' & = x_Q \cos{\varphi} - x_P \sin{\varphi} \\
y_Q' & = y_Q \cos{\varphi} - y_P \sin{\varphi} \nonumber 
\end{align}

Equations (15) and (16) are incorporated into the following C++ implementation of the \texttt{PlotEllipticArc} function: 
\begin{verbatim}
    void PlotEllipticArc(FIXED xC, FIXED yC, 
                         FIXED xP, FIXED yP, FIXED xQ, FIXED yQ, 
                         float astart, float asweep, int k)
    {
        float cosb, sinb;
        FIXED swangle;    

        xP -= xC;  
        yP -= yC;
        xQ -= xC;  
        yQ -= yC;
        if (astart != 0)
        {
            // Set new conjugate diameter end points P' and Q' 
            float cosa = cos(astart);
            float sina = sin(astart);
            FIXED x = xP*cosa + xQ*sina;
            FIXED y = yP*cosa + yQ*sina;
    
            xQ = xQ*cosa - xP*sina;
            yQ = yQ*cosa - yP*sina;
            xP = x;
            yP = y;
        }

        // If sweep angle is negative, switch direction
        if (asweep < 0)
        {
            xQ = -xQ;
            yQ = -yQ;
            asweep = -asweep;
        }
        swangle = 65536*asweep;
        EllipseCore(xC, yC, xP, yP, xQ, yQ, swangle, k);

        // Plot arc end point
        cosb = cos(asweep);
        sinb = sin(asweep);
        xP = xP*cosb + xQ*sinb;
        yP = yP*cosb + yQ*sinb;
        PlotPoint(xP+xC, yP+yC);
    }
\end{verbatim}
This function uses floating-point arithmetic to calculate the arc end point---and also the arc starting point if the specified arc starting angle is nonzero. All the points in between are calculated by the \texttt{EllipseCore} function using fixed-point arithmetic. The \texttt{cos} and \texttt{sin} functions called in this listing are declared in the standard C header file \texttt{math.h}.

On entry, the \texttt{PlotEllipticArc} function converts the conjugate diameter end points, $P$ and $Q$, to center-relative coordinates. Function parameter \texttt{astart} is the starting angle of the arc, as measured from the first conjugate diameter end point, $P$, and is positive in the direction of the second conjugate diameter end point, $Q$. Parameter \texttt{asweep} is the sweep angle, and is positive in the same direction as \texttt{astart}. Both angles are in radians. If \texttt{astart} is nonzero, the function uses equations (15) and (16) to calculate the new conjugate diameter end points, $P'$ and $Q'$, which are then passed to the \texttt{EllipseCore} function in place of $P$ and $Q$. However, a sweep angle that is negative and thus extends \emph{away} from end point $Q=(x_Q,y_Q)$ is first converted to a positive angle that extends \emph{toward} the point $\widetilde Q = (-x_Q,-y_Q)$ on the opposite end of the same conjugate diameter as $Q$. The final point plotted by \texttt{EllipseCore} typically falls short of the arc end point by a fraction of an angular increment. Before returning, the \texttt{PlotEllipticArc} function plots the arc end point.

Strictly speaking, the \texttt{PlotEllipticArc} function's \texttt{astart} and \texttt{asweep} parameters specify the arc's start and sweep angles \emph{on the unit circle, not on the ellipse}\footnote{Recall that equations (2) describe the rotation of a point on the unit circle in \mbox{Figure 2}, and that equations (3) describe the corresponding movement of the affine\:\!-transformed image of this point on the ellipse.}. It's fair to ask whether this is a convenient and intuitive way to specify an arc of an ellipse.

\begin{figure}
\centering
\includegraphics[width=300pt]{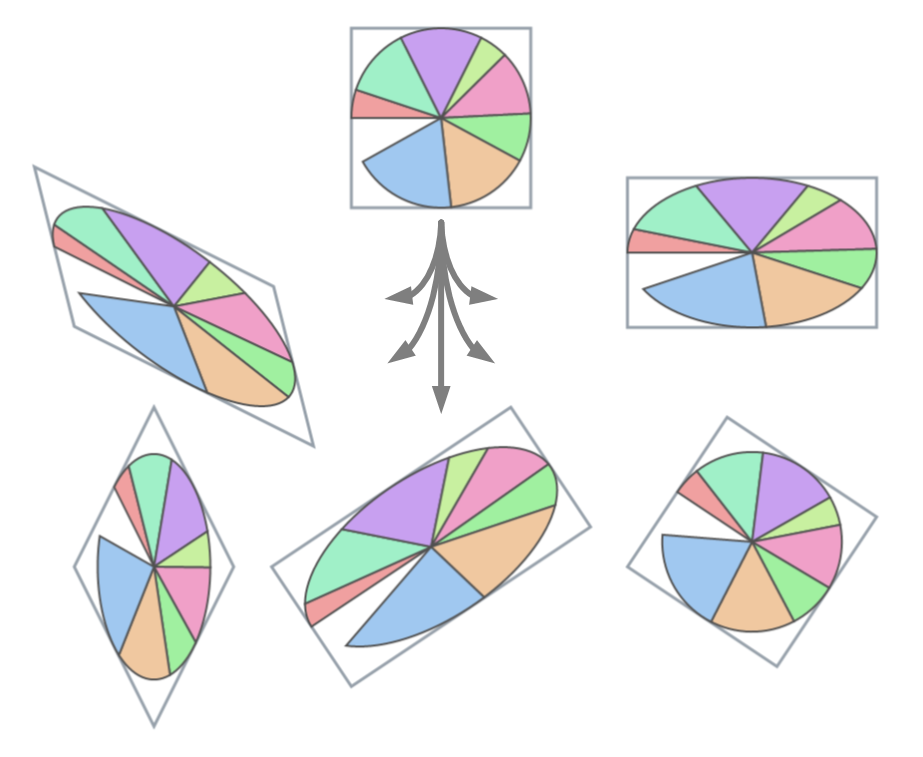}
\caption{A collection of pie charts drawn by \texttt{PlotEllipticArc} that share identical arc start and sweep angles.}
\end{figure}

To explore this issue, \mbox{Figure 3} shows a collection of pie charts---and their enclosing squares, rectangles, and parallelograms. Each arc in the figure is drawn by a \texttt{PlotEllipticArc} function call. If the circular pie chart at top center is assumed to be the original, master copy, then the others can be viewed as \mbox{3-D} images of the same pie chart as seen from different viewing angles. The arc start and sweep angles that are used to create the circular pie chart at top center are identical to those used to create the other five pie charts. The only thing that changes from one pie chart to the next is the viewing transformation, which is specified by the coordinates at the ellipse center and at the two conjugate diameter end points. Assuming that the user who constructs these six pie charts interprets them as transformed images of the same pie chart, the \texttt{PlotEllipticArc} function's interface is in accord with the user's intuition.

Arcs plotted by the \texttt{PlotEllipticArc} function share an important property with B\'ezier curves: both are \emph{affine-invariant} (Rogers and Adams [12],  Watt and Watt [13]). For a B\'ezier curve, applying any affine transformation to the vertices of the control polygon produces the same transformed image as does individually transforming the points on the curve. Similarly, for an arc constructed by the \texttt{PlotEllipticArc} function, application of any affine transformation to the ellipse center point and the two conjugate diameter end points has the same effect as individually transforming the points on the arc. 

\section{Flatness testing for ellipses} 

As previously described, the ShapeGen graphics library [6] approximates a curved shape by plotting a series of points along the curve, and then connecting each pair of adjacent points with a straight line segment---a chord. To control the spacing between points, the library user specifies a \emph{flatness} parameter, which is the maximum error tolerance, in pixels, between a chord and the curve segment that it represents. Thereafter, as the user draws various curves, the library automatically adjusts the spacing between plotted points on each curve to meet the flatness requirement.

The user can specify a smaller flatness value to improve the smoothness of the plotted curve, or can specify a larger flatness value to decrease the processing time and memory required to plot the curve.

In the previous section, the \texttt{DrawEllipse} and \texttt{DrawEllipticArc} functions require the caller to supply a parameter value \texttt{k}, which specifies the approximate angular increment ($\varepsilon = 1/2^k$ for $k = 0,1,2...$) between successive points plotted on the ellipse or elliptic arc. Thus, the caller must either calculate the correct angular increment in advance of each function call, or make a series of manual adjustments based on trial-and-error inspections of the results.

Clearly, a more convenient option is to allow the caller to specify a global flatness parameter in advance of any calls to these functions.

To make this improvement, we remove the last call parameter, \texttt{k}, from the \texttt{DrawEllipse} and \texttt{DrawEllipticArc} functions, and also from the \texttt{EllipseCore} function. The \texttt{EllipseCore} function is further modified to call a new function, which we will name \texttt{AngularInc}\cprotect\footnote{The name \verb|AngularInc| is a bit misleading because this function doesn't return angular increment $\varepsilon$. Instead, it returns $k = -\log_2 \varepsilon$. However, multiplication of a fixed-point value $x$ by $\varepsilon$ is conveniently implemented in C++ as \verb|x >> k|.}, that calculates and returns an appropriate nonnegative, integer value for the \texttt{k} parameter. The determination of this value is based on the global flatness parameter and on the size and shape of the ellipse specified by conjugate diameter end points $P$ and $Q$.

Here's the C++ declaration for the \texttt{AngularInc} function:
\begin{verbatim}
    int AngularInc(FIXED xP, FIXED yP, FIXED xQ, FIXED yQ); 
\end{verbatim}
The input parameters to this function are the center-relative $x$-$y$ coordinates of end points $P$ and $Q$, and the return value is the \texttt{k} parameter. The function determines the smallest value of \texttt{k} (and, thus, the largest angular increment) for which the chord-to\:\!-arc error does not violate the flatness requirement.

With this modification, the \texttt{EllipseCore} function plots points at regular angular increments that are specified by the \texttt{k} value returned by the \texttt{AngularInc} function.

\begin{figure}
\centering
\includegraphics[width=180pt]{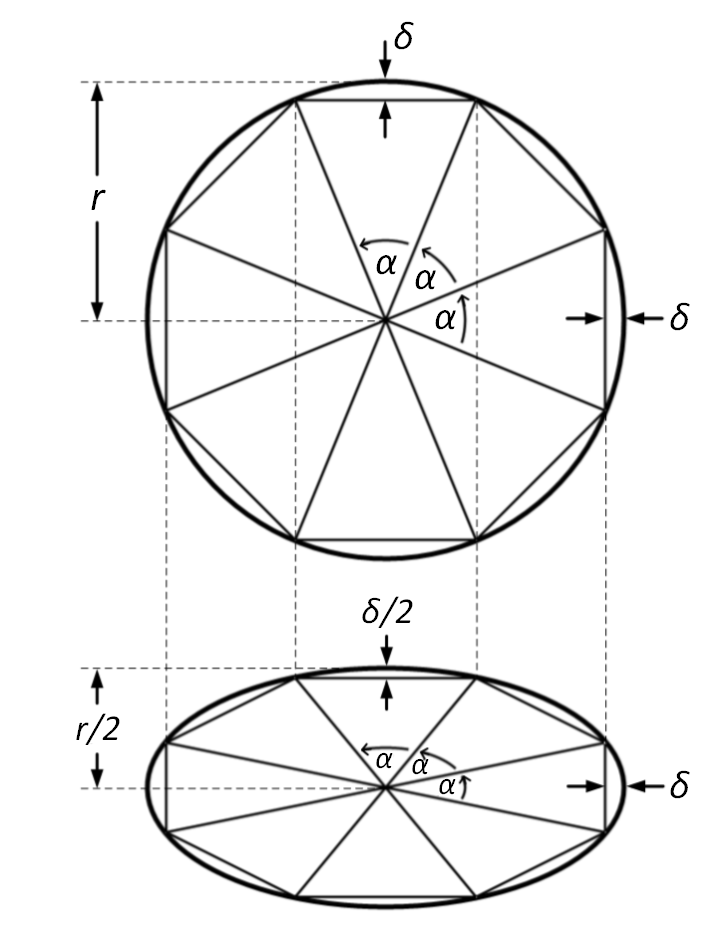}
\caption{Chord-to\:\!-arc error $\delta$ for a circle and an ellipse.}
\end{figure}

In \mbox{Figure 4}, regular angular increments result in uniformly spaced points around the circle at the top; but when the same angular increments\footnote{In the lower part of \mbox{Figure 4}, the angles labeled $\alpha$ are obviously not equal if the ellipse is viewed as a plane figure. They \emph{are} equal, however, if the ellipse is viewed as a copy of the circle above that has been rotated in three dimensions about a horizontal line through its center.} are applied to the ellipse below, the spacing varies around the ellipse. To account for the variance in the chord-to\:\!-arc error around the ellipse, the angular increment must be made small enough to accommodate the worst-case error.

For the ellipse in \mbox{Figure 4}, the largest chord-to\:\!-arc error, $\delta$, occurs at the ends of its major axis. The chord-to\:\!-arc error in the circle above is also $\delta$; this circle's diameter is equal to the length of the ellipse's major axis. We conclude that the largest chord-to\:\!-arc error for an ellipse occurs at the ends of its major axis, and that this error matches the chord-to\:\!-arc error of a circle whose diameter equals the length of the ellipse's major axis.

The \emph{auxiliary circle} (Weisstein [14]) of an ellipse is the circle whose diameter matches the length of the major axis of the ellipse, and whose center coincides with the center of the ellipse. We can refine the description of the \texttt{AngularInc} function to be that it determines the smallest \texttt{k} parameter (and, thus, the largest angular increment) that satisfies the flatness requirement \emph{for the auxiliary circle of the ellipse}.

\begin{figure}
\centering
\includegraphics[width=200pt]{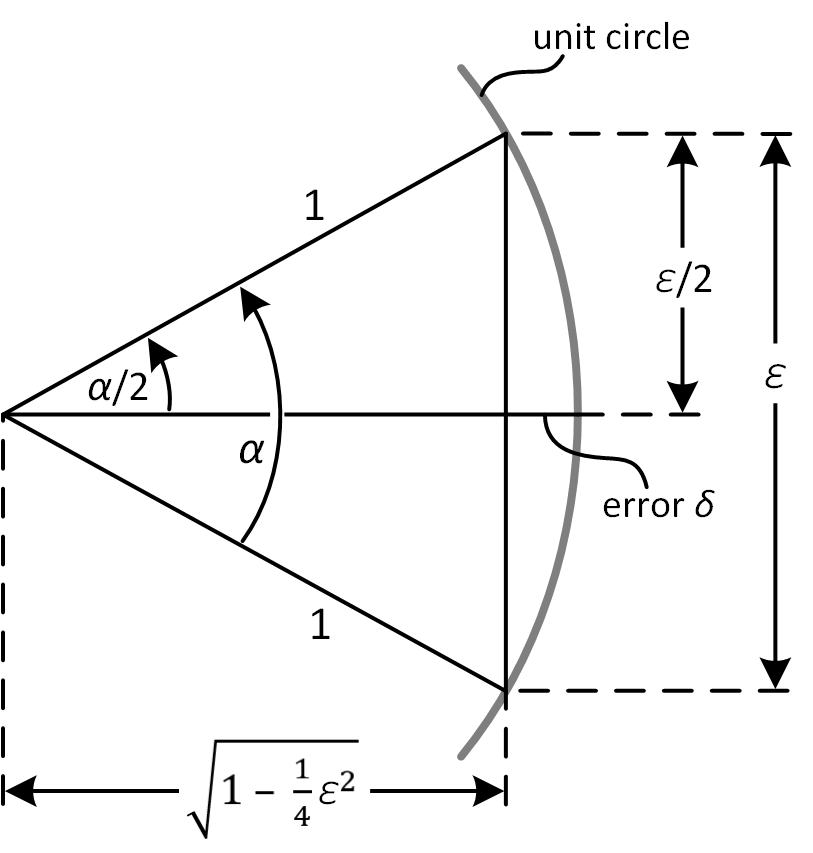}
\caption{Chord-to\:\!-arc error $\delta$ on the unit circle.}
\end{figure}

How does the chord-to\:\!-arc error for a circle depend on $\varepsilon = 1/2^k$, the approximate angular increment? \mbox{Figure 5} shows the error $\delta$ for a chord of length $\varepsilon$ on the unit circle. This figure also illustrates the relationship between $\varepsilon$ and the precise angular increment, $\alpha$, as described by equations (10). The distance from the midpoint of the chord to the circle's center is $\cos \frac{\alpha}{2} = \sqrt{1-\frac{1}{4}\varepsilon^2}$. Thus, the chord-to\:\!-arc error is $\delta = 1 - \sqrt{1-\frac{1}{4}\varepsilon^2}$. This result for the unit circle is easily generalized to an auxiliary circle of radius $r$, for which the error scales to
\begin{align*}
\delta = r \Big( 1 - \sqrt{1-\mathsmaller{\frac{1}{4}}\varepsilon^2}\,\, \Big) \textrm{, \quad where } \varepsilon = 1/2^k \textrm{ for } k = 0,1,2,... 
\end{align*}
Given a radius $r$, the \texttt{AngularInc} function might need to evaluate this expression several times while it searches for the smallest $k$ value for which $\delta$ does not exceed the user-specified flatness. To avoid the use of floating-point arithmetic, a truncated Taylor series can be used as an approximation to the square root, as was done previously in equation (14). The chord-to\:\!-arc error for the auxiliary circle can then be expressed as
\begin{align}
\delta & = r\Big(1 - \big( \,1 - \frac{1}{8}\varepsilon^2 - \frac{1}{128}\varepsilon^4 - \frac{1}{1024}\varepsilon^6 - ... \,\big) \Big) \nonumber  \\
          & = r\Big( \,\frac{1}{8}\varepsilon^2 + \frac{1}{128}\varepsilon^4 + \frac{1}{1024}\varepsilon^6 + ... \,\Big)
\end{align}
In this form, $\delta$ can be calculated using fixed-point arithmetic. All terms inside the parentheses in equation (17) are negative powers of two, and multiplications of these terms by $r$ can be performed as right-shift operations.

The following C++ implementation of the \texttt{AngularInc}  function uses equation (17) to calculate chord-to\:\!-arc error $\delta$ on an ellipse's auxiliary circle, but omits the 6th-order and higher terms in the Taylor series:
\begin{verbatim}
    int AngularInc(FIXED xP, FIXED yP, FIXED xQ, FIXED yQ)
    {
        FIXED r = AuxRadius(xP, yP, xQ, yQ);
        FIXED err2 = r >> 3;    // 2nd-order term
        FIXED err4 = r >> 7;    // 4th-order term

        for (int k = 0; k < KMAX; ++k)
        {
            if (_flatness >= err2 + err4)
                return k;

            err2 >>= 2;
            err4 >>= 4;
        }
        return KMAX;
    }
\end{verbatim}
On entry, this function obtains the radius \texttt{r} (in pixels) of the auxiliary circle by calling a function named \texttt{AuxRadius}, which will be discussed shortly. The global flatness parameter, \verb|_flatness|, is a 16.16 fixed-point value. \texttt{KMAX} is an integer constant that limits how large \texttt{k} can grow, and, thus, limits how small the angular increment can be.

How big should \texttt{KMAX} be? Well, that largely depends on the resolution of the graphics display. Let $r_{max}$ be the radius (in pixels) of the auxiliary circle for the largest ellipse the user might reasonably try to draw on the display, and let $\delta_{min}$ be the smallest flatness setting that the graphics library supports. The following approximation is obtained by truncating after the quadratic term in the Taylor series in equation (17):
\begin{align*}
\delta_{min}\, \ge \, r_{max}\Big( 1  - \sqrt{1 - \mathsmaller{\frac{1}{4}}\varepsilon^2}\,\, \Big)
\, \approx \, \frac{1}{8} \, \varepsilon^2 \, r_{max}
\end{align*}
where $\varepsilon = 1/2^{k_{max}}$, and $k_{max}$ is the value of the \texttt{KMAX} constant in the previous listing. Solving for $k_{max}$, we have
\begin{align*}
k_{max}\, \gtrsim\, \frac{1}{2} \log_2 \mathlarger( \,\frac{r_{max}}{8\, \delta_{min}} \; \mathlarger)
\end{align*}
For example, with a maximum radius of 5,000 pixels and a minimum flatness of a quarter of a pixel, we have $k_{max} \gtrsim \frac{1}{2} \log_2 ( \frac{5000}{8 \times 0.25} ) = 5.644$. And so we set \texttt{KMAX} $ = 6$.

In the preceding \texttt{AngularInc} listing, the arguments passed to the \texttt{AuxRadius} function are the center-relative coordinates of two conjugate diameter end points, $P$ and $Q$, that define an ellipse; \texttt{AuxRadius} returns the radius (in pixels) of the ellipse's auxiliary circle. Precise, closed-form solutions to the problem of finding this radius exist (Said [15], McCartin [16], Weisstein [17]). For example, the square of the radius can be calculated as [15]
\begin{align}
r^2 = \frac{1}{2}\Big(A + C + \sqrt{(A - C)^2 + B^2}\;\Big)
\end{align}
where $A = y_P^2 + y_Q^2$, $B = -2(x_P y_P + x_Q y_Q)$, and $C = x_P^2 + x_Q^2$. (The terms $A$, $B$, and $C$ in these expressions are the first three coefficients in the implicit equation for the ellipse; for more information, see \mbox{Appendix C}.) However, these solutions require floating-point arithmetic and might run slowly on less-powerful processors. A solution that uses only fixed-point addition and shift operations might therefore be preferable, especially if it quickly generates a good approximation of the radius.

To implement an \texttt{AuxRadius} function that has the desired characteristics, we can use the simple strategy shown in \mbox{Figure 6} for estimating the radius of an ellipse's auxiliary circle. The basic idea is to calculate vector lengths $|OP|$, $|OQ|$, $|OJ'|$, and $|OK'|$. We then choose the greatest of these lengths as our estimate for the radius of the ellipse's auxiliary circle. The details follow.

\begin{figure}
\centering
\includegraphics[width=300pt]{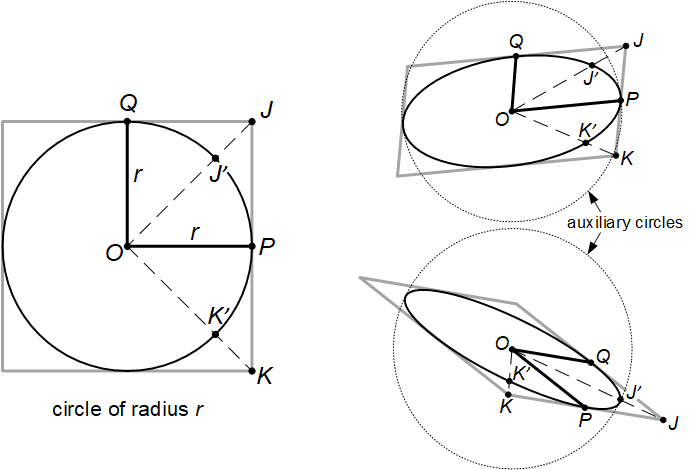}
\caption{A simple strategy for estimating the radius of an ellipse's auxiliary circle.}
\end{figure}

In Figure 6, the circle and two ellipses are centered at origin $O$. For each of these three figures, we are given end points $P = (x_P,y_P)$ and $Q = (x_Q,y_Q)$ of a pair of conjugate diameters. We use vector addition to find the corner points $J = (x_P+x_Q,y_P+y_Q)$ and $K = (x_P-x_Q,y_P-y_Q)$ of the enclosing square or parallelogram. For the circle on the left, each of the following four lengths equals radius $r$: $|OP|$, $|OQ|$, $|OJ'| = |OJ|/\sqrt{2}$, and $|OK'| = |OK|/\sqrt{2}$. For an ellipse that's nearly circular, any one of these four lengths is still a reasonably good approximation to the radius of the auxiliary circle.

For the two eccentric ellipses on the right side of \mbox{Figure 6}, however, some of the corresponding lengths serve as better approximations to the auxiliary circle radius than others. On the top right, point $P$ comes closest to touching the auxiliary circle, and thus $r \approx |OP|$ gives the best approximation. On the bottom right, point $J'$ comes closest to touching the auxiliary circle, and so $r \approx |OJ'|$ gives the best approximation.

To determine the points $J'$ and $K'$ for the ellipses on the right side of \mbox{Figure 6}, we exploit a useful property of affine transformations: they preserve ratios between directed segments that lie on the same line. If we view these two ellipses as affine\:\!-transformed images of the circle on the left, we see that the ratios $J' : J$ and $K' : K$ must be preserved. Thus, we have $|OJ'| = |OJ|/\sqrt{2}$ and $|OK'| = |OK|/\sqrt{2}$ for the two ellipses as well as for the circle.

An \texttt{AuxRadius} function that uses the strategy just described requires a means of estimating vector lengths $|OP|$, $|OQ|$, $|OJ|$, and $|OK|$ in \mbox{Figure 6}. The following C++ function takes as input parameters the $x$ and $y$ components of a vector, and returns the approximate length of the vector:
 
\begin{verbatim}
    FIXED VLen(FIXED x, FIXED y)
    {
        x = abs(x);
        y = abs(y);
        if (x > y)
            return x + max(y/8, y/2 - x/8);

        return y + max(x/8, x/2 - y/8);
    }
\end{verbatim}
This function uses fixed-point addition and shift\footnote{The assumption here is that the compiler converts divisions by constants that are powers of two into right-shift operations.} operations to estimate the vector length. The \texttt{abs} function call in this listing returns the absolute value of the input argument. The \texttt{max} function call returns the greater of the two argument values. Assume that \texttt{abs} and \texttt{max} are both inline functions.

The \texttt{VLen} function is largely based on a method set forth in an ancient Tamil poem (Hattangadi [18]) for estimating the length of a hypotenuse. Given a right triangle with sides of length $x$ and $y$, where $0 \le y \le x$, this method estimates the length of the hypotenuse as $x + y/2 - x/8$. For $y < x/3$, however, a much better approximation is $x + y/8$. With this enhancement, the error in the \texttt{VLen} function's return value falls within the range $-2.8$ to $+0.78$ percent.

The following C++ implementation of the \texttt{AuxRadius} function follows the strategy presented in the discussion of \mbox{Figure 6} to estimate the radius of an ellipse's auxiliary circle:
\begin{verbatim}
    FIXED AuxRadius(FIXED xP, FIXED yP, FIXED xQ, FIXED yQ)
    {
        FIXED dP = VLen(xP, yP);
        FIXED dQ = VLen(xQ, yQ);
        FIXED dJ = VLen(xP + xQ, yP + yQ);
        FIXED dK = VLen(xP - xQ, yP - yQ);
        FIXED r1 = max(dP, dQ);
        FIXED r2 = max(dJ, dK); 
    
        return max(r1 + r1/16, r2 - r2/4);
    }
\end{verbatim}
The input parameters to this function are the center-relative coordinates of two conjugate diameter end points that define the ellipse. The \texttt{VLen} function is called to obtain the approximate lengths of the four vectors $OP$, $OQ$, $OJ$, and $OK$ shown in \mbox{Figure 6}. The longest of the vectors $OP$, $OQ$, $OJ'$, and $OK'$ typically falls just short of reaching the auxiliary circle, and \texttt{VLen} tends to slightly underestimate the vector lengths it calculates. The \texttt{AuxRadius} function compensates for these shortfalls by slightly inflating the radius lengths that it calculates. The larger of the two lengths $|OP|$ and $|OQ|$ is inflated by \mbox{$100 \!\times \!(1/16)$} percent, and the larger of the two lengths $|OJ'|$ and $|OK'|$ is inflated by $100 \!\times \!(\frac{3}{4}\sqrt{2} - 1)$ percent.

The error in the \texttt{AuxRadius} function's return value falls within the range $-4.2$ to $+7.1$ percent. This level of accuracy should provide sufficient flatness control for many graphics applications.

On faster processors, however, using equation (18) might work as well as or better than the strategy shown in  \mbox{Figure 6}. Also, the approximate value returned by the \texttt{AuxRadius} function above might not suffice for a graphics library that treats a user's flatness setting as a strict requirement. To keep the performance issues in perspective, recall that \texttt{AuxRadius} is called just once during the initialization phase of an \texttt{EllipseCore} function call.

\section{Minsky circle variants}

It's not too surprising that the Minsky circle-generating algorithm has been invented more than once, given its simplicity. About the same time that Marvin Minsky came up with the algorithm, it was independently developed by David Mapes, working on a PDP-1 at Lawrence Livermore National Laboratory\footnote{https://www.computer-history.info/Page2.dir/pages/Mapes.html}. Mapes says his colleague George Michael ``made me aware that I had rediscovered what was discovered long ago.''

Mapes experimented with several variants of the circle algorithm's inner loop. Two variants drew (approximate) circles, but in different directions. Two other variants drew hyperbolic arcs. These variants reveal interesting properties of the generalized algorithm that might lead to useful applications.

The inner loop of Minsky's circle algorithm was presented earlier in equations (6) and (7), and is repeated here, except that $x$-$y$ (rather than $u$-$v$) coordinates are used:\\[-20pt]
\begin{align*}
x_n & = x_{n-1} - \varepsilon \!\: y_{n-1}    \\
y_n & = y_{n-1} + \varepsilon \!\: x_n 
\end{align*}
where $0 < \varepsilon \le 1$. This inner loop plots points in the \emph{forward} direction; that is, in the direction of increasing angles. For example, if the starting point $(x_0,y_0)$ lies on the $+x$-axis, subsequent points are plotted in the direction of the $+y$-axis.

Given the same starting point, $(x_0,y_0)$, the following variant plots points on the same approximate circle as the preceding inner loop, but steps in the \emph{reverse} direction:\\[-20pt]
\begin{align*}
y_n & =  y_{n-1} - \varepsilon \!\: x_{n-1}   \\
x_n & = x_{n-1} + \varepsilon \!\: y_n
\end{align*}
The reader can verify that if the first (\emph{forward} direction) of these two inner loops is executed once, and then the second (\emph{reverse} direction) is executed once, the ending coordinates exactly match the starting coordinates. (Assume the $\varepsilon$ values are the same.) By extension, if the first loop is executed a million times, followed by a million iterations of the second loop, the ending point must still equal the starting point.

This next variant draws a hyperbolic arc:
\begin{align*}
x_n & = x_{n-1} + \varepsilon \!\: y_{n-1}    \\
y_n & = y_{n-1} + \varepsilon \!\: x_n 
\end{align*}
As was the case with the Minsky circle algorithm, the resulting arc is somewhat askew but approaches the ideal curve as $\varepsilon$ decreases in size. The realm of hyperbolic angles is a \emph{Through-the-Looking-Glass} version of the circular-angle realm, and the error analysis of this variant has a few quirks but otherwise closely parallels the analysis of the circle algorithm given in \mbox{Appendix A}. A precise, error-corrected hyperbolic arc algorithm can therefore be constructed that is nearly identical to the ellipse algorithm presented in this paper. The only changes required are sign changes for some terms (from plus to minus, and vice versa), plus the replacement of the circular sine and cosine values in the \texttt{PlotEllipticArc} function with hyperbolic sine and cosine values in the corresponding function, \texttt{PlotHyperbolicArc}.

Another similarity between ellipses and hyperbolas is that both have conjugate diameters. Just as any ellipse can be specified by a center point and two conjugate diameter end points, so can any hyperbola (and its conjugate hyperbola).

Here's a final variant:\\[-20pt]
\begin{align*}
y_n & =  y_{n-1} - \varepsilon \!\: x_{n-1}   \\
x_n & = x_{n-1} - \varepsilon \!\: y_n
\end{align*}
Given the same starting point as the preceding inner loop, this variant plots points on the same hyperbola, but steps in the reverse direction.

In typical graphics applications, hyperbolic arcs are not used as extensively as circular and elliptic arcs, but are useful nonetheless. Users of graphics drawing programs might have difficulty using other types of curves to convincingly fake hyperbolic arcs.

Also, a fast hyperbolic arc algorithm could have uses beyond the drawing of curves. For example, the rendering of a radial gradient requires a per-pixel calculation of the distance from the center of the radial pattern, and this calculation is typically done with a square\:\!-root operation. However, a hyperbolic arc is formed by the intersection of a radial gradient cone with a plane perpendicular to the display surface. If this arc is constructed by an algorithm that uses fixed-point shifts and adds to plot radial distances along a scan line, most of the square\:\!-root operations can be eliminated.

\section{Conclusion}

Although parametric equations are typically best-suited for plotting points on curves, the straightforward implementation of the parametric equations for circles and ellipses would require a graphics engine to evaluate sine and cosine terms at each point. To avoid the computational cost of such evaluations, a number of alternative methods have been developed to more cheaply approximate circular and elliptic arcs (Blinn [11], Ri\v{s}kus [19]). For example, PostScript interpreters construct approximate circular arcs in piecewise fashion from cubic B\'ezier curves (Adobe [8] p. 365).

This paper has presented a parametric ellipse algorithm that plots points on a circular or elliptic arc that precisely match those that would be plotted by explicitly calculating sine and cosine values at each point. Yet the computational cost per point plotted by the ellipse algorithm is a mere six integer additions and four shift operations.

Arguably, the simplest, most intuitive way for a graphics user to control the shape and orientation of an ellipse is by specifying the square, rectangle, or parallelogram in which the ellipse is inscribed. The ellipse algorithm uses an equivalent set of parameters to specify an ellipse---namely, a center point, and the end points, $P$ and $Q$, of a pair of conjugate diameters of the ellipse. A parallelogram and its inscribed ellipse share a common center point, and $P$ and $Q$ are simply the midpoints of two adjacent sides of the parallelogram.

Like B\'ezier curves, elliptic arcs plotted by the ellipse algorithm are affine\:\!-invariant. Thus, applying an affine transformation to the three points---center, $P$, and $Q$---that define the ellipse that contains the arc will generate the same transformed image as individually transforming the points on the arc.

Finally, some graphics applications require control over the direction in which points are plotted on a curve---for example, when constructing the boundary of a shape to be filled according to the nonzero\:\!-winding-number rule. The ellipse algorithm provides straightforward direction control. For a full ellipse, it always starts at point $P$ and plots points in the direction of point $Q$. When plotting an elliptic arc, the arc's starting angle is specified relative to $P$ and is positive in the direction of $Q$. The arc's sweep angle is positive in the same direction. Because the ellipse algorithm relies on these simple conventions instead of using direction flags, it is immune to problems caused by affine transformations that produce reflections.

\section{References}

\begin{enumerate}
\item Van\,Aken, J., Simar, R. (1988). ``A Conic Spline Algorithm,'' \emph{TMS34010 Application Guide}, Texas Instruments Incorporated, 255-278.

\item Van\,Aken, J., Simar, R. (1992). ``A Parametric Elliptical Arc Algorithm,'' \emph{Graphics Gems III}, Academic Press, Inc., 164-172.

\item Beeler, M., Gosper, R.W., Schroppel, R. (1972). ``Item 149 (Minsky),'' \emph{HAKMEM}, Massachusetts Institute of Technology Artificial Intelligence Laboratory AIM-239. \texttt{https://dspace.mit.edu/handle/1721.1/6086}

\item Paeth, A.W. (1990). ``A Fast Algorithm for General Raster Rotation,'' \emph{Graphics Gems}, Academic Press, Inc., 179-195.

\item Ziegler Hunts, C., Ziegler Hunts, J., Gosper, R.W., Holloway, J. (2011). \emph{Minskys \& Trinskys: Exploring an Early Computer Algorithm}, 3rd ed., ZH2G\&H (self-published through Blurb, Inc.).\\
\verb|https://www.blurb.com/b/2172660-minskys-trinskys-3rd-edition|

\item Van\,Aken, J. (2020). ``ShapeGen: A Lightweight, Open-Source 2-D Graphics Library Written in C++,'' ResearchGate.\\
\verb|https://www.researchgate.net/profile/Jerry_Van_Aken| 

\item Van\,Aken, J. (2018). ``A Rotated Ellipse from Three Points,'' ResearchGate. 
\verb|https://www.researchgate.net/profile/Jerry_Van_Aken|

\item \emph{PostScript Language Reference Manual}, 3rd ed. (1999). Adobe Systems Incorporated.

\item Foley, J.D., van Dam, A., Feiner, S.K., Hughes, J.F. (1990). \emph{Computer Graphics: Principles and Practice}, 2nd ed., Addison-Wesley Publishing Company, 952-953.

\item Newman, W.M., Sproull, R.F. (1979). \emph{Principles of Interactive Computer Graphics}, McGraw-Hill, Inc., 27-28.

\item Blinn, J. (1996). ``How Many Ways Can You Draw a Circle?,'' \emph{A Trip Down the Graphics Pipeline}, Morgan Kaufmann Publishers, Inc., 4-5.

\item Rogers, D.F., Hart, J.A (1990). \emph{Mathematical Elements for Computer Graphics}, 2nd ed., McGraw-Hill, Inc., 291-307.

\item Watt, A., Watt, M. (1992). \emph{Advanced Animation and Rendering Techniques}, Addison-Wesley, 69.

\item Weisstein, E.W. ``Auxiliary Circle.'' From MathWorld---A Wolfram Web Resource. \texttt{https://mathworld.wolfram.com/AuxiliaryCircle.html}

\item Said, M.A. (2003). ``Calibration of an Ellipse's Algebraic Equation and Direct Determination of its Parameters,'' \emph{Acta Mathematica Academiae Paedagogicae Ny\'iregyh\'aziensis}, \textbf{19}, 221-225.

\item McCartin, B.J. (2013). ``A Matrix Analytic Approach to Conjugate Diameters of an Ellipse,'' \emph{Applied Mathematical Sciences}, \textbf{7}(36), 1797-1810.

\item Weisstein, E.W. ``Ellipse.'' From MathWorld---A Wolfram Web Resource. \texttt{https://mathworld.wolfram.com/Ellipse.html}

\item Hattangadi, A.A. (2002). \emph{Explorations in Mathematics}, Sangan Books Ltd, 59-61.

\item Ri\v{s}kus, A. (2006). ``Approximation of a Cubic B\'ezier Curve by Circular Arcs and Vice Versa,'' \emph{Information Technology and Control}, \textbf{35}(4).

\end{enumerate}


\appendices
\noindent

\setcounter{equation}{0}
\renewcommand{\theequation}{\thesection.\arabic{equation}}

\section{\mbox{Analysis of the circle generator}}
This appendix analyzes the error in Marvin Minsky's circle-generating algorithm\footnote{Ray Simar's original analysis of the circle generator appeared as an appendix in [1], which was published as part of an application guide that is now both out of print and unavailable on the Web. I have reproduced the substance of Simar's analysis here, but edited it to fit into the context of the current paper, and split the analysis into two separate appendices, A and B, to improve continuity.}. Equations (8) and (9) in the main text are closed-form expressions for the state of the circle generator's $u$ and $v$ variables after  $n$ iterations, where $n = 0,1,2,...$ The following analysis will derive these two equations from the difference equations in the circle generator's inner loop.

The inner loop consists of equations (6) and (7) in the main text, which are repeated here, except that $x$-$y$ (rather than $u$-$v$) coordinates are used:
\begin{align}
x_n & = x_{n-1} - \varepsilon \, y_{n-1}    \\
y_n & = \varepsilon \, x_n + y_{n-1} 
\end{align}
where $0 < \varepsilon \le 1$, $(x_0,y_0)$ is the initial point on the circle, and $n=1,2,...$

These equations can be expressed in matrix form by substituting the right-hand side of equation (A.1) in place of $x_n$ in equation (A.2) to yield
\begin{align*}
\begin{bmatrix}
x_n \\ y_n
\end{bmatrix}
& =
\begin{bmatrix}
1 & -\varepsilon \\
\varepsilon & 1-\varepsilon^2
\end{bmatrix}
\begin{bmatrix}
x_{n-1} \\ y_{n-1}
\end{bmatrix}
\end{align*}
For the sake of brevity, this matrix equation can be expressed as $\mathbf{x}_n = \mathbf{M} \mathbf{x}_{n-1}$, where\\[-20pt]   
\begin{align*}
\mathbf{x}_n = 
\begin{bmatrix}
x_n \\ y_n
\end{bmatrix}
\qquad \textrm{and} \qquad \mathbf{M} =  
\begin{bmatrix}
1 & -\varepsilon \\
\varepsilon & 1-\varepsilon^2
\end{bmatrix}
\end{align*}
Our goal is to obtain a closed-form solution for the matrix equation $\mathbf{x}_n = \mathbf{M}^n \mathbf{x}_0$, which we expect will be equivalent to equations (8) and (9) in the main text.

The eigenvalues $\lambda_1$ and $\lambda_2$ of matrix $\mathbf{M}$ are the roots of the \emph{characteristic equation} 
$\textrm{det}(\mathbf{M} - \lambda \mathbf{I}) = 0$, which expands to the following:
\begin{align*}
0 &= \textrm{det}\! 
\begin{bmatrix*}
1 - \lambda & -\varepsilon \\[+2pt]
\varepsilon & 1-\varepsilon^2 - \lambda \,
\end{bmatrix*} \\[+5pt]
  &= \lambda^2 + (\varepsilon^2 - 2)\lambda + 1
\end{align*}
The quadratic formula yields the following two solutions for $\lambda$:
\begin{align*}
\lambda = 1 - \mathsmaller{\frac{1}{2}} \varepsilon^2 \: \pm \, \varepsilon \:\! \sqrt{\mathsmaller{\frac{1}{4}}\varepsilon^2 - 1}
\end{align*}
For the range $0 < \varepsilon < 2$, the result of the square root operation above is an imaginary number, and the preceding expression can be rewritten as
\begin{align}
\lambda = 1 - \mathsmaller{\frac{1}{2}} \varepsilon^2 \: \pm \, i \:\! \varepsilon \:\! \sqrt{1 - \mathsmaller{\frac{1}{4}}\varepsilon^2}
\end{align}
where $i^2 = -1$. Thus, the two solutions for $\lambda$ are complex numbers, which can be written in the form\\[-18pt]
\begin{align*}
\lambda &= \rho \:\! e^{\pm i \alpha}
\end{align*}
where $\lambda$ is a vector on the complex plane, $\rho$ is the vector's length, and angle $\alpha$ specifies vector's orientation. Length $\rho$ is calculated by applying the Pythagorean theorem to equation (A.3), as follows:
\begin{align*}
\rho &= \sqrt{ (1 - \mathsmaller{\frac{1}{2}} \varepsilon^2 )^2 + \varepsilon^2 (1 - \mathsmaller{\frac{1}{4}}\varepsilon^2)} \\[+2pt]
  &= 1
\end{align*}
Thus, the two solutions for $\lambda$ are unit vectors and can be expressed as
\begin{align}
\lambda &= e^{\pm i \alpha} \\
  &= \cos \alpha \pm i \sin \alpha \nonumber
\end{align}
By comparing this last expression with equation (A.3), we see that
\begin{align}
\cos \alpha &= 1 - \mathsmaller{\frac{1}{2}} \varepsilon^2 \\[+2pt] 
\sin \alpha &= \varepsilon \, \sqrt{1 - \mathsmaller{\frac{1}{4}}\varepsilon^2}
\end{align}

As shown in \mbox{Appendix B}, the equation for $\mathbf{M}^n$ has the form
\begin{align}
\mathbf{M}^n &= a_n \mathbf{M} + b_n \mathbf{I} 
\end{align}
where the values of coefficients $a_n$ and $b_n$ are yet to be determined. Similarly, the equations for the $n$th powers of eigenvalues $\lambda_1$ and $\lambda_2$ have the form
\begin{align}
     \lambda_1^n &= a_n \lambda_1 + b_n \\[+4pt]
     \lambda_2^n &= a_n \lambda_2 + b_n
\end{align}
The two solutions for $\lambda$ in equation (A.4) are $\lambda_1 = e^{+ i \alpha}$ and $\lambda_2 = e^{- i \alpha}$. The values $e^{+i \alpha}$ and $e^{-i \alpha}$ can therefore be substituted for $\lambda_1$ and $\lambda_2$ in equations (A.8) and (A.9) to yield \\[-18pt]
\begin{align*}
     e^{+ i n \alpha} &= a_n e^{+ i \alpha} + b_n \\
     e^{- i n \alpha} &= a_n e^{- i \alpha} + b_n
\end{align*}
When these two simultaneous equations are solved for coefficients $a_n$ and $b_n$, their values are found to be\footnote{The solution uses the identities $\sin \theta = \mathlarger{\frac{e^{i \theta}-e^{-i \theta}}{2 i}}$ and $\cos \theta = \mathlarger{\frac{e^{i \theta}+e^{-i \theta}}{2}}$.} \\[-18pt]
\begin{align*}
     a_n &= \frac{\sin n \alpha }{\sin \alpha} \\
     b_n &= \cos n \alpha - \frac{\sin n \alpha }{\sin \alpha} \cos \alpha
\end{align*}
Equation (A.7) can now be expanded as follows:
\begin{align*}
    \mathbf{M}^n &= a_n \mathbf{M} + b_n \mathbf{I} \\
                         &=
\begin{bmatrix*}
  \, a_n + b_n  &  - \varepsilon\, a_n \, \\[+4pt]
  \, \varepsilon\, a_n  &  a_n(1 - \varepsilon^2) + b_n \,
\end{bmatrix*} \\[+4pt]
                         &=
\begin{bmatrix*}
  \,\mathsmaller{\cos n \alpha + (1 - \cos \alpha)} \frac{\sin n \alpha}{\sin \alpha}  &
  - \varepsilon \frac{\sin n \alpha}{\sin \alpha} \\[+4pt]
  \varepsilon \frac{\sin n \alpha}{\sin \alpha}  &  
  \mathsmaller{\cos n \alpha + (1 - \varepsilon^2 - \cos \alpha)} \frac{\sin n \alpha}{\sin \alpha}\,
\end{bmatrix*}
\end{align*}
To cross-check this result, the reader can verify that $\mathbf{M}^0 = \mathbf{I}$ and $\mathbf{M}^1 = \mathbf{M}$. 

We are now ready to evaluate the matrix equation $\mathbf{x}_n = \mathbf{M}^n \mathbf{x}_0$, which yields the following values for $x_n$ and $y_n$:
\begin{align*}
  x_n &= x_0 \Big( \!\cos n \alpha + \big( 1 - \cos \alpha \big) \frac{\sin n \alpha}{\sin \alpha}\, \Big) 
            - y_0 \Big( \varepsilon \, \frac{\sin n \alpha}{\sin \alpha}\, \Big) \\[+4pt]
  y_n &= x_0 \Big( \varepsilon \, \frac{\sin n \alpha}{\sin \alpha}\, \Big) 
            + y_0 \Big( \!\cos n \alpha + \big( 1 - \varepsilon^2 - \cos \alpha \big) \frac{\sin n \alpha}{\sin \alpha}\, \Big) 
\end{align*}
Finally, we replace $\cos \alpha$ and $\sin \alpha$ in these two equations with the values in equations (A.5) and (A.6), and consolidate terms, to yield
\begin{align}
x_n &= x_0 \cos n \alpha - \left[ \frac{y_0 - \frac{\mathlarger{\varepsilon}}{2} x_0}{\sqrt{1 - \frac{1}{4} \varepsilon^2 }} \right] \sin n \alpha 
\\
y_n &= \left[ \frac{x_0 - \frac{\mathlarger{\varepsilon}}{2} y_0}{\sqrt{1 - \frac{1}{4} \varepsilon^2 }} \right] \sin n \alpha + y_0 \cos n \alpha
\end{align}
where $0 < \varepsilon < 2$ and $n = 0,1,2,...$ These equations are equivalent to equations (8) and (9) in the main text, except that they use $x$-$y$ (instead of $u$-$v$) coordinates. Also, typical graphics applications restrict $\varepsilon$ to the range $0 < \varepsilon \le 1$.

For an ideal circle generator, the terms in brackets in equations (A.10) and (A.11) should equal $y_0$ and $x_0$, respectively. However, the terms in brackets can be made to approach the desired values, $y_0$ and $x_0$, by decreasing the size of $\varepsilon$.

\setcounter{figure}{0} \renewcommand{\thefigure}{A.\arabic{figure}} 

\begin{figure}
\centering
\includegraphics[width=200pt]{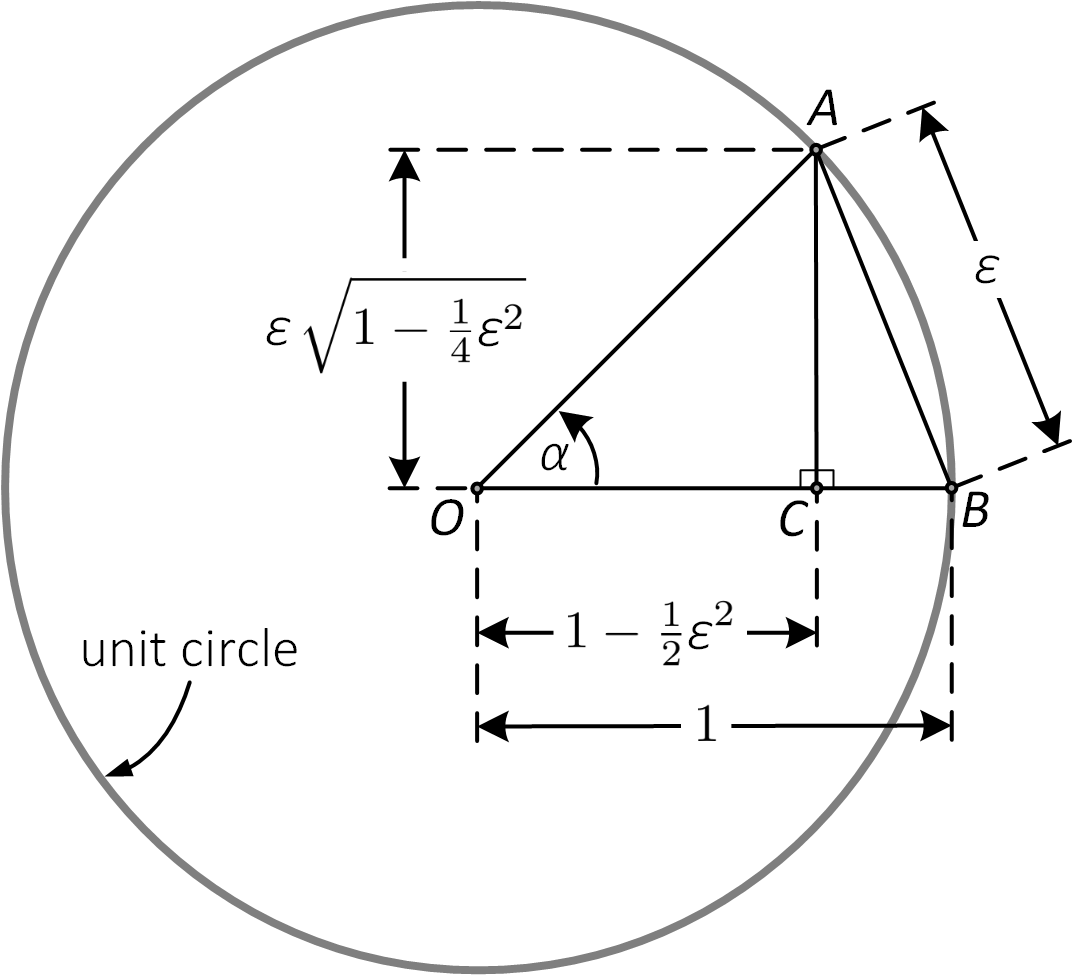}
\caption{If chord $AB$ on the unit circle subtends an angle $\alpha$ for which \mbox{$\sin \alpha = \varepsilon \sqrt{1-\frac{1}{4}\varepsilon^2}$} and $\cos \alpha = 1-\frac{1}{2} \varepsilon^2$, the length of the chord is $\varepsilon$.}
\end{figure} 

Decreasing $\varepsilon$ also decreases angle $\alpha$ because the two are related by equations (A.5) and (A.6). In fact, $\alpha \rightarrow \varepsilon$ as $\varepsilon \rightarrow 0$. \mbox{Figure A.1} shows that if a chord $AB$ subtends an angle $\alpha$ on the unit circle, and if $\sin \alpha = \varepsilon \sqrt{1-\frac{1}{4}\varepsilon^2}$ and $\cos \alpha = 1-\frac{1}{2} \varepsilon^2$, the length of the chord is $\varepsilon$. This fact follows from the observation that chord $AB$ is the hypotenuse of right triangle $ABC$, and the sum of the squares of sides $AC$ and $BC$ of this triangle equals the square of $AB$.

Applying the \emph{half-angle formulas} to equation (A.5) yields the relations
\begin{align*}
\sin \frac{\alpha}{2} = \frac{\varepsilon}{2} \qquad \textrm{and} \qquad
\cos \frac{\alpha}{2} = \sqrt{1 - \mathsmaller{\frac{1}{4}} \varepsilon^2 }
\end{align*}
These two expressions, together with \mbox{Figure 5} in the main text, are helpful for further understanding the relationship between $\varepsilon$ and $\alpha$. In particular, this figure shows how the chord-to\:\!-arc error on the unit circle depends on chord length $\varepsilon$.

\setcounter{equation}{0}

\section{The \emph{n}th power of a $\mathbf{2\mathbf{\times}2}$ matrix}

A $2 \! \times \! 2$ matrix $\mathbf{M}$ has two eigenvalues, $\lambda_1$ and $\lambda_2$.
This appendix will show that the $n$th power of $\mathbf{M}$ can be expressed in an equation of the form
\begin{align}
\mathbf{M}^n = a_n \mathbf{M} + b_n \mathbf{I} \qquad n = 0,1,2,...
\end{align}
where $\mathbf{I}$ is the $2 \!\times\! 2$ identity matrix, and the values of coefficients $a_n$ and $b_n$ are functions of the four coefficients $m_{ij}$ of $\mathbf{M}$. Also shown here is that the $n$th powers of eigenvalues $\lambda_1$ and $\lambda_2$ can be expressed in similar form; that is, as
\begin{align}
\lambda_1^n = a_n \lambda_1 + b_n \\ 
\lambda_2^n = a_n \lambda_2 + b_n
\end{align}
where coefficients $a_n$ and $b_n$ are identical to those in the expression for $\mathbf{M}^n$. For the case $\lambda_1 \ne \lambda_2$, we can solve equations (B.2) and (B.3) for $a_n$ and $b_n$ in terms of $\lambda_1$, $\lambda_2$, $\lambda_1^n$, and $\lambda_2^n$. Coefficients $a_n$ and $b_n$ can then be used to provide a closed-form solution for $\mathbf{M}^n$. These results are used in \mbox{Appendix A}.
 
We start with the following $2 \!\times\! 2$ matrix:
\begin{align*}
\mathbf{M} =
\begin{bmatrix}
m_{11}  & m_{12}  \\ m_{21}  & m_{22}  
\end{bmatrix} 
\end{align*}
The eigenvalues $\lambda_1$ and $\lambda_2$ of $\mathbf{M}$ are the roots of the \emph{characteristic equation} $\textrm{det}(\mathbf{M} - \lambda \mathbf{I}) = 0$, which we expand as follows:
\begin{align*}
0 &= \textrm{det}\! \left[ \,
\begin{matrix*}[l]
m_{11} - \lambda  & \, m_{12}  \\ m_{21}  & \,  m_{22} - \lambda 
\end{matrix*} \, \right] \\[+5pt]
  &= \lambda^2 - (m_{11} + m_{22})\lambda^1 + (m_{11} m_{22} - m_{12} m_{21}) \lambda^0
\end{align*}

The \emph{Cayley-Hamilton theorem} states that every square matrix satisfies its own characteristic equation. Applying this theorem to the characteristic equation for $\mathbf{M}$ yields
\begin{align*}
0 &= \mathbf{M}^2 - (m_{11} + m_{22}) \mathbf{M}^1 + (m_{11} m_{22} - m_{12} m_{21}) \mathbf{M}^0
\end{align*} 
which can be rearranged as follows:
\begin{align*}
\mathbf{M}^2 = (m_{11} + m_{22}) \mathbf{M}^1 + (m_{12} m_{21} - m_{11} m_{22}) \mathbf{M}^0 
\end{align*}
For convenience, we define $a_2 = m_{11} + m_{22}$ and $b_2 = m_{12} m_{21} - m_{11} m_{22}$. We also note that $\mathbf{M}^0 =\mathbf{I}$, so that the preceding expression for $\mathbf{M}^2$ simplifies to 
\begin{align*}
\mathbf{M}^2 = a_2 \mathbf{M} + b_2 \mathbf{I} 
\end{align*}
With this expression for $\mathbf{M}^2$ serving as the basis, expressions for successively higher powers of $\mathbf{M}$ can be generated recursively as follows:
\begin{align*}
\mathbf{M}^3 &= \mathbf{M} ( \mathbf{M}^2 ) \\
  &= \mathbf{M} ( a_2 \mathbf{M} + b_2 \mathbf{I} ) \\
  &= a_2 \mathbf{M}^2 + b_2 \mathbf{M} \\
  &= a_2 ( a_2 \mathbf{M} + b_2 \mathbf{I} ) + b_2 \mathbf{M} \\
  &= ( a_2^2 + b_2 ) \mathbf{M} + a_2 b_2 \mathbf{I} \\
  &= a_3 \mathbf{M} + b_3 \mathbf{I}\;\! , 
\quad \textrm{where \:} a_3 = a_2^2 + b_2 \textrm{\: and \:} b_3 = a_2 b_2
\end{align*} 
\begin{align*}
\mathbf{M}^4 &= \mathbf{M} ( \mathbf{M}^3 ) \\
  &= \mathbf{M} ( a_3 \mathbf{M} + b_3 \mathbf{I} ) \\
  &= a_3 \mathbf{M}^2 + b_3 \mathbf{M} \\
  &= a_3 ( a_2 \mathbf{M} + b_2 \mathbf{I} ) + b_3 \mathbf{M} \\
  &= ( a_2 a_3 + b_3 ) \mathbf{M} + a_3 b_2 \mathbf{I} \\
  &= a_4 \mathbf{M} + b_4 \mathbf{I}\;\! , 
\quad \textrm{where \:} a_4 = a_2 a_3 + b_3 \textrm{\: and \:} b_4 = a_3 b_2 \\[-3pt]
&\vdots
\end{align*} \\[-36pt] 
\begin{align*}
\phantom{|||||||||||||}
\mathbf{M}^n &= \mathbf{M} ( \mathbf{M}^{n-1} ) \\
  &= \mathbf{M} ( a_{n-1} \mathbf{M} + b_{n-1} \mathbf{I} ) \\
  &= a_{n-1} \mathbf{M}^2 + b_{n-1} \mathbf{M} \\
  &= a_{n-1} ( a_2 \mathbf{M} + b_2 \mathbf{I} ) + b_{n-1} \mathbf{M} \\
  &= ( a_2 a_{n-1} + b_{n-1} ) \mathbf{M} + a_{n-1} b_2 \mathbf{I} \\
  &= a_n \mathbf{M} + b_n \mathbf{I}\;\! , 
\quad \textrm{where \:} a_n = a_2 a_{n-1} + b_{n-1} \textrm{\: and \:} b_n = a_{n-1} b_2
\end{align*}

Similar methods can be used to derive a sequence of equations for $\lambda^2,\lambda^3,...,\lambda^n$. First, to obtain $\lambda^2$, the characteristic equation previously given for $2 \!\times\! 2$ matrix $\mathbf{M}$ can be rearranged as follows:
\begin{align*}
\lambda^2 = a_2 \lambda + b_2 
\end{align*}
where coefficients $a_2$ and $b_2$ are defined as before. With this expression for $\lambda^2$ serving as the basis, expressions for successively higher powers of $\lambda$ can be generated recursively as follows:
\begin{align*}
\lambda^3 &= \lambda (\lambda^2) = a_3 \lambda + b_3
\\ 
\lambda^4 &= \lambda (\lambda^3) = a_4 \lambda + b_4
\\[-5pt] &\vdots \\
\lambda^n &= \lambda (\lambda^{n-1}) = a_n \lambda + b_n
\end{align*}
where the coefficient values $a_i$ and $b_i$, for $i = 3, 4,...,n$, are identical to those in the preceding expressions for $\mathbf{M}^3, \mathbf{M}^4,..., \mathbf{M}^n$. The expression $\lambda^n = a_n \lambda + b_n$ just above is valid for the two eigenvalues of $\mathbf{M}$, $\lambda_1$ and $\lambda_2$. 

Thus far, we have confirmed that the recurrent relations for $\mathbf{M}^n$, $\lambda_1^n$, and $\lambda_2^n$ have the form shown in equations (B.1), (B.2), and (B.3). The final step is to derive closed-form expressions for coefficients $a_n$ and $b_n$.  

The two simultaneous equations $\lambda_1^n = a_n \lambda_1 + b_n$ and $\lambda_2^n = a_n \lambda_2 + b_n$ can be solved to obtain the following closed-form expressions for $a_n$ and $b_n$, which are valid for $\lambda_1 \ne \lambda_2$:  \\[-16pt] 
\begin{align*}
a_n &= \frac{\lambda_1^n - \lambda_2^n}{\lambda_1 - \lambda_2} \\[+6pt]
b_n &= \frac{\lambda_1^n + \lambda_2^n}{2} - \frac{\lambda_1^n - \lambda_2^n}{2} \big(\frac{\lambda_1 + \lambda_2}{\lambda_1 - \lambda_2}\big)
\end{align*}
With these two expressions, we can immediately write $\mathbf{M}^n = a_n \mathbf{M} + b_n \mathbf{I}$ as a closed-form expression. The reader can verify that $\mathbf{M}^0 = \mathbf{I}$ and $\mathbf{M}^1 = \mathbf{M}$, so that our expression for $\mathbf{M}^n$ is valid for integer $n \ge 0$.

\setcounter{equation}{0}

\section{The implicit equation of an ellipse}

This appendix derives the implicit equation for an ellipse from the parametric ellipse equations. The ellipse is centered at the $x$-$y$ origin and is specified by conjugate diameter end points $P = (x_P,y_P)$ and $Q = (x_Q,y_Q)$. The derivation of $P$ and $Q$ from the implicit equation for the ellipse is also discussed.

\subsection{Deriving the implicit equation of an ellipse from the parametric equations}

Equation (18) in the main text is the formula for the square of radius $r$ of an ellipse's auxiliary circle (Said [15]), and is repeated here:
\begin{align}
r^2 = \frac{1}{2}\Big(A + C + \sqrt{(A - C)^2 + B^2}\;\Big)
\end{align}
The terms $A$, $B$, and $C$ in this equation are the first three coefficients in the implicit equation for the ellipse, which has the form
\begin{align*}
f(x,y) &= A x^2 + B x y + C y^2 + D x + E y + F = 0 
\end{align*}
How do we determine the coefficient values $A$,$B$,...,$F$ for an ellipse defined by conjugate diameter end points $P$ and $Q$? We will consider the case of an ellipse centered at the $x$-$y$ origin, for which $D = E = 0$.

Equations (3) in the main text are parametric equations that define an origin-centered ellipse in terms of the end points, $P=(x_P,y_P)$ and $Q=(x_Q,y_Q)$, of a pair of the ellipse's conjugate diameters. The equations are repeated here:
\begin{align*}
x(\theta) & = x_P \cos{\theta} + x_Q \sin{\theta} \\[+2pt]
y(\theta) & = y_P \cos{\theta} + y_Q \sin{\theta}
\end{align*}
where $0 \le \theta \le 2\pi$. These two simultaneous equations can be solved for the two unknowns $\sin{\theta}$ and $\cos{\theta}$, to yield
\begin{align*}
\sin{\theta} &= \frac{x_P y(\theta) - x(\theta) y_P}{x_P y_Q - x_Q y_P} \\[+4pt]
\cos{\theta} &= \frac{x_Q y(\theta) - x(\theta) y_Q}{x_Q y_P - x_P y_Q}
\end{align*}
The implicit equation for the ellipse is obtained by substituting the two expressions above into the identity $\sin^2{\theta} + \cos^2{\theta} = 1$, as follows:
\begin{align*}
1 &= \Big( \,\frac{x_P\!\: y - x\!\: y_P}{x_P\!\: y_Q - x_Q\!\: y_P}\, \Big)^2 + \; \Big( \,\frac{x_Q\!\: y - x\!\: y_Q}{x_Q\!\: y_P - x_P\!\: y_Q}\, \Big)^2 \\[+8pt]
&= \frac{(y_P^2 + y_Q^2) x^2 - 2(x_P\!\: y_P + x_Q\!\: y_Q) x\!\: y + (x_P^2 + x_Q^2) y^2}
            {(x_P\!\: y_Q - x_Q\!\: y_P)^2}
\end{align*}
The terms can then be rearranged to produce an equation of the form
\begin{equation}
f(x,y) = Ax^2 + Bxy + Cy^2 + Dx + Ey + F = 0   \\
\end{equation}
\begin{align*}
\\[-22pt]
\text{where} \quad A & = y_P^2 + y_Q^2   \\
B & = -2(x_P\!\: y_P + x_Q\!\: y_Q)   \\
C & = x_P^2 + x_Q^2   \\
D & = E = 0   \\
F & = -(x_P\!\: y_Q - x_Q\!\: y_P)^2
\end{align*}

\subsection{Deriving parametric equations of an ellipse from the implicit equation}

If we are given coefficients $A$-$F$ of the implicit equation for an ellipse, as in equation (C.2), we can derive a pair of conjugate diameter end points, $P=(x_P,y_P)$ and $Q=(x_Q,y_Q)$, that describe the same ellipse. (The implicit equation represents an ellipse if $B^2-4AC<0$.) If the center of this ellipse does not lie at the $x$-$y$ origin, the derivation of points $P$ and $Q$ will be simplified if we first translate the ellipse's center to the origin before proceeding further.

If the coefficients $D$ and $E$ of the implicit equation are nonzero, the ellipse is centered at some point $(x_0,y_0)$ other than the origin. To translate the center of this ellipse to the origin, the implicit equation can be transformed as follows:
\begin{align*}
0 &= f(x+x_0,y+y_0) \\
   &= A(x+x_0)^2 + B(x+x_0)(y+y_0) + C(y+y_0)^2 \\
   & \qquad \qquad + D(x+x_0)+ E(y+y_0) + F  \\
   &= A x^2 + B x y + C y^2 + (2Ax_0 + By_0 + D)x + (2Cy_0 + Bx_0 + E)y \\
   & \qquad \qquad + (Ax_0^2 + Bx_0y_0 + Cy_0^2 + Dx_0 + Ey_0 + F) \\
   &= A'x^2 + B'x y + C'y^2 + D'x + E'y + F'
\end{align*}
where coefficients $A'$-$F'$ define the new, origin-centered ellipse; these coefficients are specified, as follows, in terms of the coefficients, $A$-$F$, of the original ellipse:
\begin{align*}
&A' = A, \quad B' = B, \quad C' = C, \\
&D' = 2Ax_0 + By_0 + D, \quad E' = 2Cy_0 + Bx_0 + E, \\
&F' = Ax_0^2 + Bx_0y_0 + Cy_0^2 + Dx_0 + Ey_0 + F
\end{align*}
For an origin-centered ellipse, coefficients $D'$ and $E'$ must be zero. The two simultaneous equations $D'=0$ and $E'=0$ (that is, $2Ax_0 + By_0 + D = 0$ and $2Cy_0 + Bx_0 + E = 0$) can be solved for the two unknowns $x_0$ and $y_0$, to yield
\begin{align*}
x_0 = \frac{BE-2CD}{4AC-B^2} \quad \textrm{and} \quad y_0 = \frac{BD-2AE}{4AC-B^2}
\end{align*}
The resulting $x_0$ and $y_0$ values can then be used to calculate $F'$.

After the implicit equation for an ellipse has been translated to the origin, equation (C.2) describes how the coefficients\footnote{Going forward, we'll use $A$-$F$ rather than $A'$-$F'$ to denote the coefficients in the origin-centered ellipse equation.} $A$-$F$ are defined in terms of a pair of conjugate diameter end points, $P=(x_P,y_P)$ and $Q=(x_Q,y_Q)$, of the ellipse. However, our problem here is the inverse of that discussed in the previous subsection. That is, given coefficients $A$-$F$ of the implicit equation for an ellipse, we want to find a pair of conjugate diameter end points $P$ and $Q$ that describe the same ellipse. Deriving $P$ and $Q$ from $A$-$F$ is complicated by the fact that an ellipse has an infinite number of possible diameters, each of which has a conjugate. Thus, there are an infinite number of conjugate diameter end points, $P$ and $Q$, that all describe the same ellipse. To find a solution for a pair of conjugate diameters, we must specify that they are in some way unique. For example, a typical choice is to set $P$ and $Q$ to end points of the major and minor axes of the ellipse, which are also conjugate diameters [7,\,15].

If, however, we are looking for a derivation that is simple and straightforward, a good choice is to align one diameter with either the $x$- or $y$-axis. For example, we can place point $P$ on the $x$-axis by setting $y_P=0$. Consequently, at point $P$, any term in the implicit equation that contains $y$ goes to zero. Equation (C.2) then simplifies to $f(x_P,y_P)=Ax_P^2 + F = 0$, which has two solutions for $x_P$; for example, we could arbitrarily choose $x_P=+\sqrt{-F/A}$. The $y_P^2$ term in the expression for coefficient $A$ in equation (C.2) also goes to zero, and we are left with $A=y_Q^2$, which has two solutions for $y_Q$; for example, we could arbitrarily choose $y_Q=+\sqrt{A}$. Finally, the term $x_Py_P$ in the expression for coefficient $B$ goes to zero, so that $B=-2x_Qy_Q$, and $x_Q$ has the unique solution \mbox{$x_Q=-\frac{B}{2y_Q}$}, where $y_Q$ has the value previously chosen. The two resulting conjugate diameter end points, $P=(x_P,y_P)$ and $Q=(x_Q,y_Q)$, describe the same ellipse as coefficients $A$-$F$ in the implicit equation.

\subsection{Calibrating an implicit equation}

If coefficients $A$-$F$ in equation (C.2) are all multiplied by the same value $\eta \ne 0$, the resulting implicit equation still represents the same ellipse. But a subtle problem has arisen: the value $y_Q=+\sqrt{A}$ calculated in the preceding paragraph has now been multiplied by $\sqrt{\eta}$ and, thus, the value of $y_Q$ changes even though the altered equation describes the same ellipse as before. A similar problem occurs with the auxiliary circle radius $r$ derived from equation (C.1). We need a way to avoid this type of error.

Said [15] points out that there is a unique scaling of the coefficients \mbox{$A$-$F$}, specified by a \emph{calibration number} $\delta$, that preserves the relationship between these coefficients and invariant properties of the ellipse, such as radius $r$ in equation (C.1). For an origin-centered ellipse, the calibration number is calculated as
\begin{align*}
\delta = \frac{-4F}{4AC - B^2}
\end{align*}
An implicit ellipse equation whose coefficients $A$-$F$ have all been multiplied by the calibration number is said to be \emph{calibrated}. Of course, there's no need to multiply the coefficients by $\delta$ if the equation is already calibrated (that is, if $\delta = 1$).

To verify that equation (C.2) is already calibrated, the calibration number for this equation is calculated, as follows, from coefficients $A$-$F$:
\begin{align*}
\delta &= \frac{4(x_P\!\: y_Q -x_Q\!\: y_P)^2}{4(y_P^2 + y_Q^2)(x_P^2 + x_Q^2) - 4(x_P\!\: y_P + x_Q\!\: y_Q)^2} \\
          &= \frac{x_P^2\!\: y_Q^2 - 2\!\: x_P\!\: y_P\!\: x_Q\!\: y_Q + x_Q^2\!\: y_P^2}
                      {x_P^2\!\: y_Q^2 - 2\!\: x_P\!\: y_P\!\: x_Q\!\: y_Q + x_Q^2\!\: y_P^2} \\
          &= 1
\end{align*}
Thus, equation (C.2) is already calibrated, and the coefficient values defined for this equation can be used without modification in expressions such as equation (C.1).

\end{document}